\RequirePackage{fix-cm}
\documentclass[smallextended]{svjour3}       
\smartqed  
\usepackage{graphicx}
\usepackage{caption}
\usepackage{textcomp}
\usepackage{subfigure}
\usepackage{float}

\begin{document}

\title{Lifetime measurement of the 5d$^2$D$_{5/2}$ state in Ba$^+$}


\author{Amita Mohanty, Elwin A. Dijck,   \\
Mayerlin Nu\~{n}ez Portela, Nivedya Valappol, Andrew T. Grier,\\
Thomas Meijknecht, Lorenz Willmann, Klaus Jungmann.   
}

\institute{Amita Mohanty  \at
              Van Swinderen Institute, FMNS, University of Groningen, \\
              9747AA Groningen, The Netherlands. \\
              Tel.: +31-644747323\\
              \email{a.mohanty@rug.nl, mohanty.amita18@gmail.com}           
}
\date{Received: date / Accepted: date}

\titlerunning{Lifetime measurement of 5d$^2$D$_{5/2}$ state in Ba$^+$ }
\authorrunning{Amita Mohanty et al.} 
\maketitle

\begin{abstract}
The lifetime of the metastable 5d$^2$D$_{5/2}$ state has been measured for a single trapped Ba$^+$ ion in a Paul trap
in Ultra High Vacuum (UHV) in the 10$^{-10}$ mbar pressure range. A total of 5046 individual periods when the ion was 
shelved in this state have been recorded. A preliminary value $\tau_{D_{5/2}} = 26.4(1.7)$~s is obtained through 
extrapolation to zero residual gas pressure. 

\end{abstract}
\section{Introduction}
\label{intro}
The accurate determination of the transition probability of transitions in heavy 
alkali earth systems is an important step in the research program to measure Atomic Parity Violation 
(APV) in such systems \cite{Portela,Wansbeek3,Willmann,Portela2,Sahoo1,Sahoo2,Sahoo3,Geetha,Roberts}. In the research reported here, 
a single trapped  Ba$^+$ ion has been investigated and the  lifetime of its 5d$^2$D$_{5/2}$ state has been measured. 
This provides essential input for testing atomic structure and, in particular, the atomic wavefunctions of the involved states 
at percent level accuracy. Such  measurements are highly sensitive to variations of parameters that determine the experiment's 
performance during long periods (i.e. several hours) and which may cause systematic uncertainties. In particular, such effects 
may arise from  interactions of the ion with background gas. 

There are two main reasons for choosing single trapped Ba$^+$ ion in UHV to perform precise lifetime measurements. 
Firstly, Barium~(Ba) is a heavy alkaline earth metal. The Ba$^+$ ion has a rather simple electronic configuration. Precise 
measurements provide for accurate tests of the atomic wavefunctions. Secondly, systematic errors 
due to collisions with other particles (such as different species) are highly suppressed. 

The lifetime of the metastable 5d$^2$D$_{5/2}$ state in Ba$^+$ has been measured earlier in different experiments 
\cite{Nagourney,Madej,Plumelle,Royen,Gurell,Auchter,Yu}. Calculations are presently performed by different independent theory groups 
\cite{Sahoo1,Sahoo2,Sahoo4,Dzuba,Iskrenova,Guet,Guet1}. All the measurements to date as well as calculated values for the lifetime of 
5d$^2$D$_{5/2}$ state in Ba$^+$ have been compiled in Table 1.
\begin{table}[h]
\centering
\caption{Calculations and measurements of the lifetime of the 5d$^2$D$_{5/2}$ state in Ba$^+$(see also Fig. 6). 
Note, some of the values have been reported without error bars.}
\label{tab:1}       
\hspace*{-2cm}
\begin{tabular}{llllll}
\hspace{1.0cm}Theory & \hspace{-4.0cm}Experiments \\
\begin{tabular}{llllll}
\hline\noalign{\smallskip}
Value[s] & Year & Reference & Value[s] & Year & Reference  \\
\noalign{\smallskip}\hline\noalign{\smallskip}
29.8(3) & 2012 & \hspace*{0.45cm}\cite{Sahoo1,Sahoo2} & 31.2(0.9) & 2014 & \hspace*{0.45cm}\cite{Auchter} \\
30.3(4) & 2008 & \hspace*{0.45cm}\cite{Iskrenova} & 32.0(2.9) & 2007 & \hspace*{0.45cm}\cite{Royen} \\
30.8 & 2007 & \hspace*{0.45cm}\cite{Guet1} & 32.3 & 1997 & \hspace*{0.45cm}\cite{Yu} \\
31.6 & 2007 & \hspace*{0.45cm}\cite{Gurell} & 34.5(3.5) & 1990 & \hspace*{0.45cm}\cite{Madej} \\
30.3 & 2001 & \hspace*{0.45cm}\cite{Dzuba} & 32(5) & 1986 & \hspace*{0.45cm}\cite{Nagourney} \\
37.2 & 1991 & \hspace*{0.45cm}\cite{Guet} & 47(16) & 1980 & \hspace*{0.45cm}\cite{Plumelle} \\
\noalign{\smallskip}\hline\noalign{\smallskip}
\end{tabular}
\end{tabular}
\end{table}
\vspace*{-1.50cm}
\section{Experimental setup}
\label{sec:1}
The trap for Ba$^+$ in this experiment is a hyperbolic Paul trap \cite{Paul}. It consists of a ring electrode and two end caps
made of copper. The electrodes are mounted on a Macor holder. The chosen geometry results in a harmonic pseudopotential at the 
center of the trap when AC voltages are applied between the ring and the two endcaps. The latter are grounded. The operating RF 
frequency for the trap is $\Omega_{RF}$ = 5.44~MHz. 
The trap with its Macor holder is mounted on Oxygen Free High Conductivity (OFHC) copper base plate. In order to trap ions,
there is a Ba oven (0.9 mm diameter $\times$ 40 mm length resistively heated stainless steel tube) which contains a mixture of 
BaCO$_{3}$ and Zr. This oven produces a flux of order of $10^6$ thermal Ba atoms/s.
\begin{figure}[h]
 \begin{minipage}[t]{0.50\linewidth}
  \includegraphics[width=\textwidth]{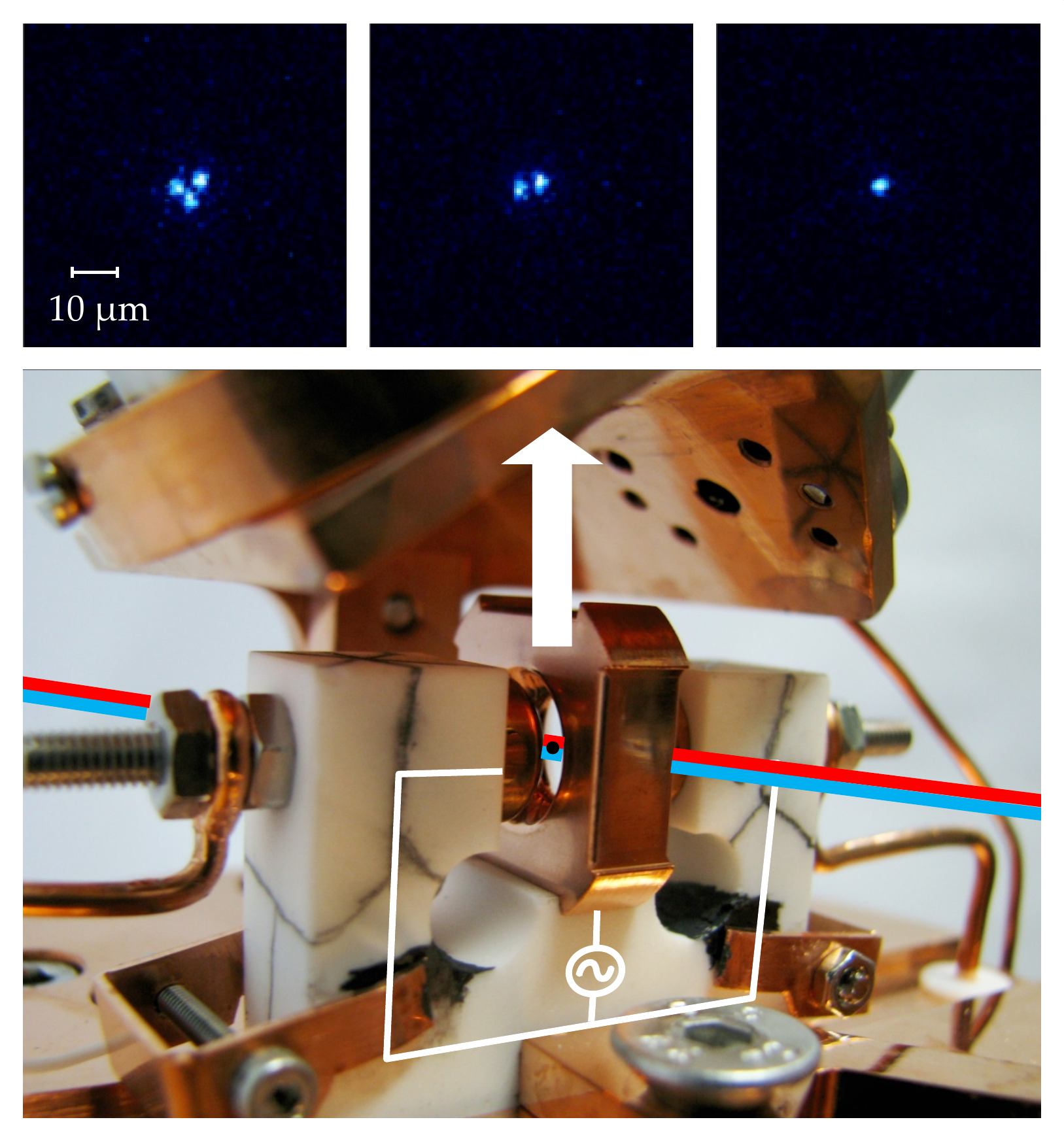}
\caption{Hyperbolic Paul trap of Ba$^+$ ion. On top, images of 3, 2 and 1 ion are given.}
\label{fig:minipage1}       
\end{minipage}
\quad
 \begin{minipage}[t]{0.50\linewidth}
  \includegraphics[width=\textwidth]{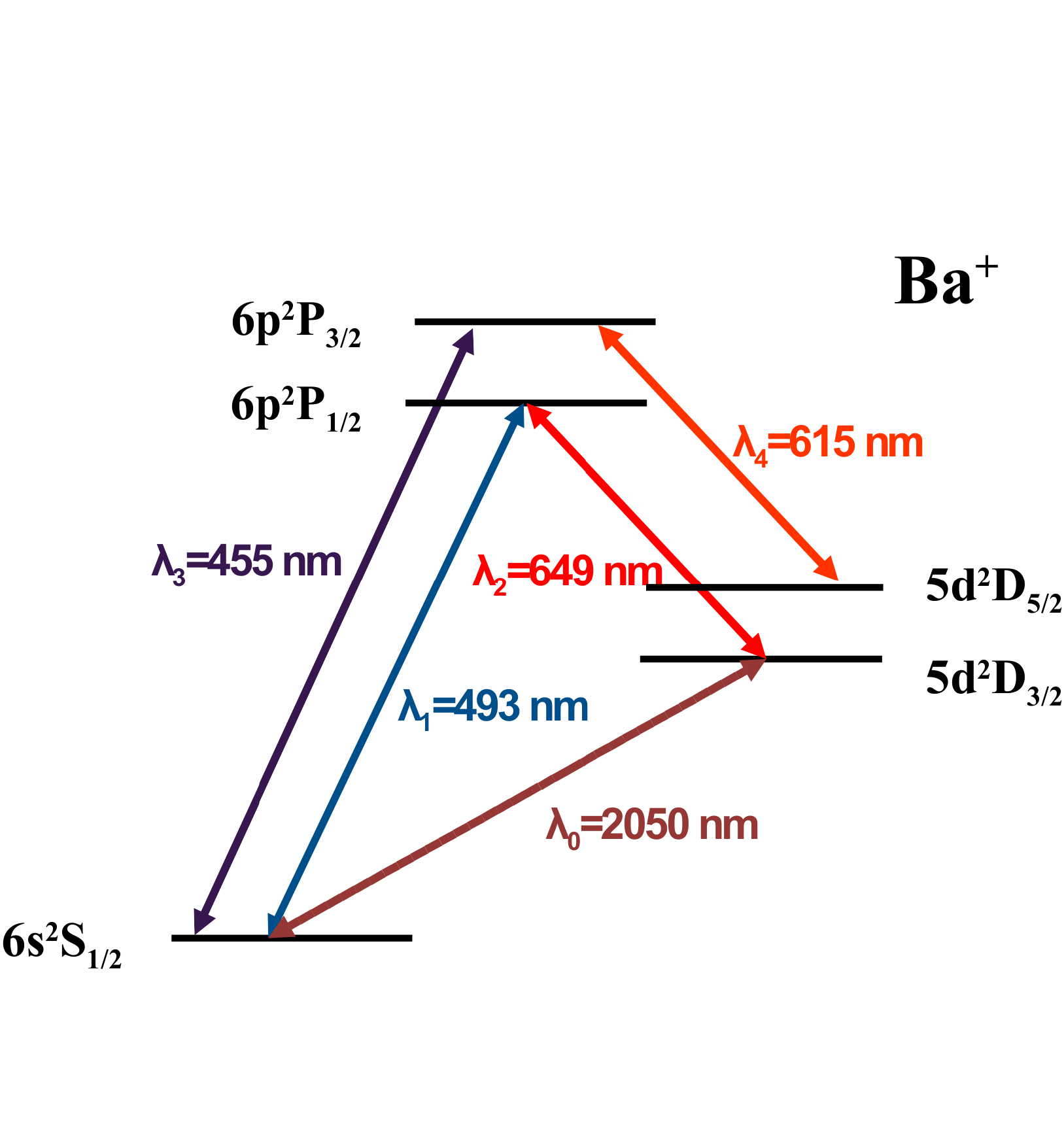}
 \caption{Level scheme of Ba$^+$ ion. The lowest $^2$S$_{1/2}$, $^2$P$_{1/2}$ and $^2$D$_{3/2}$ electronic 
 states form a closed three level system. }
 \label{fig:minipage2}
 \end{minipage}
\end{figure}
 A laser at wavelength 413~nm is used to produce Ba$^+$ ions in the trap by 
 two-photon photoionisation. We use laser light at $\lambda_{1} = 493$~nm (frequency doubled from a Coherent MBR-110 Ti:Sa 
 laser) for driving the 6s$^2$S$_{1/2}$-6p$^2$P$_{1/2}$ cooling transition and laser light at  $\lambda_{2} = 649$~nm (produced from 
 Coherent CR-699 ring dye laser) for the 6p$^2$P$_{1/2}$-5d$^2$D$_{3/2}$
 repump transition (see Fig. 2). In the experiments reported here, the power of $\lambda_{1}$ is between 6~\textmu{}W and 50~\textmu{}W 
 and that of $\lambda_{2}$ is between 6~\textmu{}W and 45~\textmu{}W. The Gaussian radius of the laser beams is about 60~\textmu{}m 
 at the position of the ion for all the measurements. Fluorescence from the 6s$^2$S$_{1/2}$-6p$^2$P$_{1/2}$ transition 
 in the Ba$^+$ ion is detected with a photomultiplier tube (PMT) and an EMCCD camera. Fig. 1 shows our hyperbolic 
 Paul trap together with the image of ions that are trapped and localized at the potential minimum of the trap.
\vspace*{-0.50cm}
 \section{Electron shelving technique}
\label{sec:2}
Ba$^+$ ions have a closed three-level system. One of the excited states, the 5d$^2$D$_{5/2}$
state, is long-lived (see Fig. 2). Simultaneous laser radiation at $\lambda_{1}$ and $\lambda_{2}$ is therefore needed 
to cool the ion in the center of the trap. When the ion is exposed to the light of two laser beams at wavelengths 
$\lambda_{1}$ and $\lambda_{2}$ (see Fig. 2), there is a closed cycle of 6s$^2$S$_{1/2}$-6p$^2$P$_{1/2}$-5d$^2$D$_{3/2}$ 
transitions. Observing the fluorescence from the 6p$^2$P$_{1/2}$-6s$^2$S$_{1/2}$ transition implies that the ion is ``not shelved'' 
in the 5d$^2$D$_{5/2}$ state. The electron shelving technique is employed in our experiment to determine the lifetime 
of the 5d$^2$D$_{5/2}$ state. With an additional fiber-coupled high power LED (M455F1) at $\lambda_{3} = 455$~nm wavelength the 
ion can be ``shelved'' to the 5d$^2$D$_{5/2}$ state via excitation to the 6p$^2$P$_{3/2}$ state and this
state's subsequent decay. The direct observation of ``quantum jumps'' in a single Ba$^+$ ion between the 5d$^2$D$_{5/2}$ 
and 6s$^2$S$_{1/2}$ states has been first demonstrated by Nagourney et al. \cite{Nagourney}.
The decay of the 6p$^2$P$_{3/2}$ state is the start of a shelving period which ends with a quantum jump from the 
5d$^2$D$_{5/2}$ state to 6s$^2$S$_{1/2}$ state. Fig.~3 displays the highest PMT count rate (2200cnts/s) when the ion is not 
shelved and the lowest count rate (600cnts/s) as background when it is shelved to the metastable 5d$^2$D$_{5/2}$ state. 
The ``on/off'' and ``off/on'' transitions in the fluorescence signal corresponds to the start and end of one single interval, when the 
ion was in the 5d$^2$D$_{5/2}$ state.
\begin{figure}[h]
  \centering
  \includegraphics[width=\textwidth]{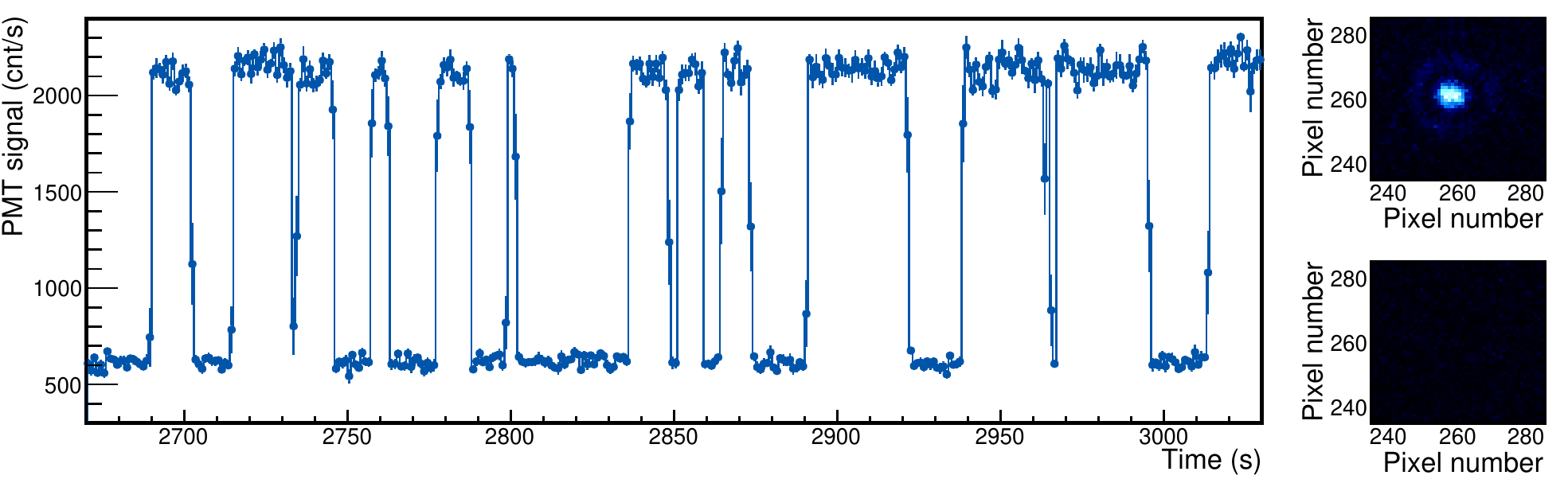}
 \caption{Quantum Jumps observed in single Ba$^+$ ion. Left : PMT count rate as a function of time. 
 Right: EMCCD image of the ion in the unshelved state(top) and shelved state(bottom).}
  \end{figure}  
  \vspace*{-0.50cm}
\section{Measurements}
\label{sec:3}
In order to measure the lifetime $\tau_{D_{5/2}}$, a total of 5046 individual shelved periods
have been recorded in 71 data samples and analysed. They were taken under in part significantly different conditions to 
enable observing and correcting for systematic errors \cite{Portela3}. Fig.~4 represents one example of the analysed
samples. It shows an exponential decay. Such a decay function is fitted to each data set using a binned 
log-likelihood method. The lifetime $\tau_{D_{5/2}}$ is obtained for each data sample from the corresponding fit parameters. 
We note that experimental situations can be created where ion heating results in longer measured durations of 
individual dark periods than the actual dwell time of the ion in the D$_{5/2}$ state. This can be seen in the slow 
recovery of the fluorescence light.
\begin{figure}[h]
\begin{minipage}[t]{0.48\linewidth}
 \includegraphics[width=\textwidth]{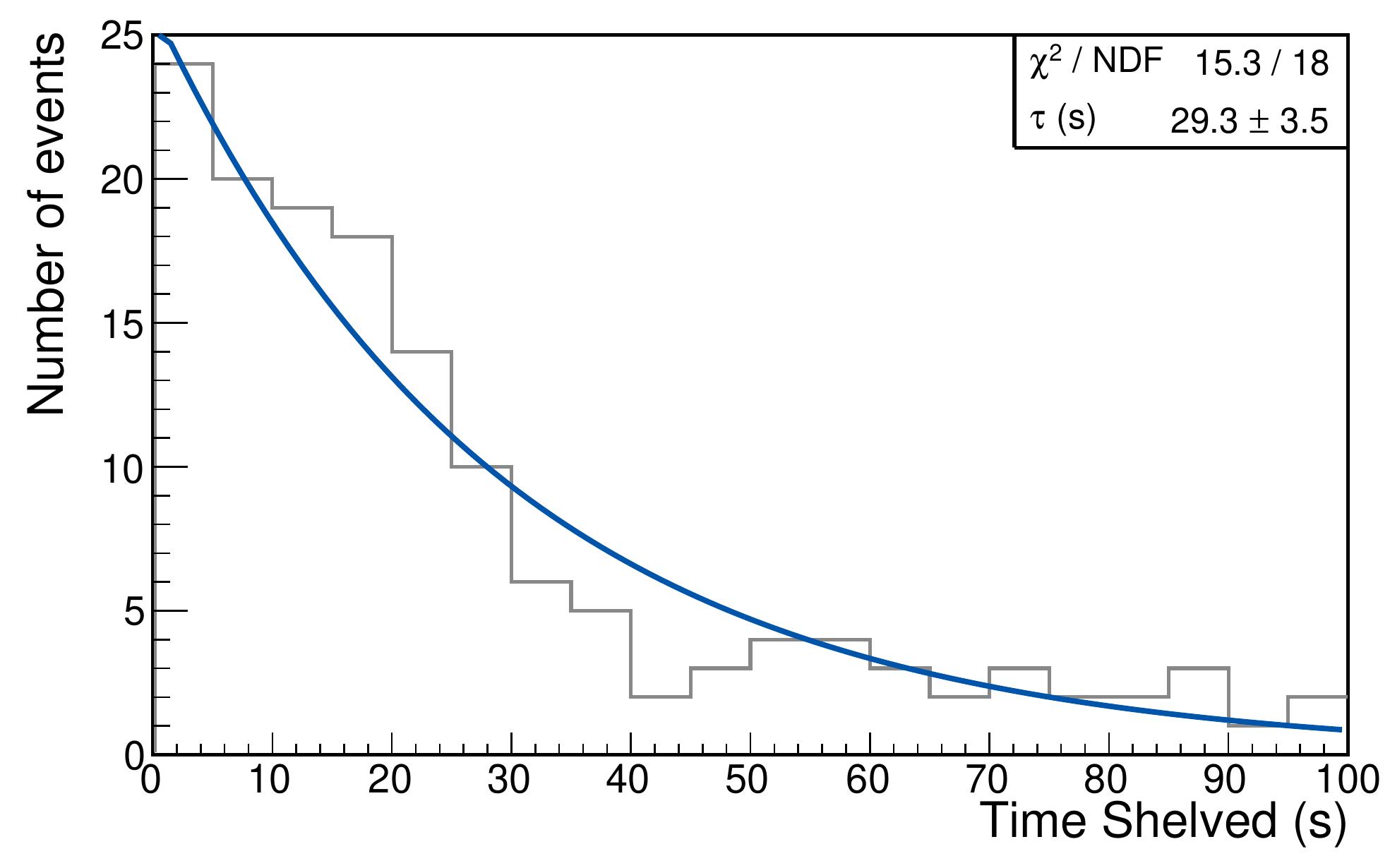}
\caption{One sample of the lifetime measurements in single Ba$^+$ ion with 96 shelved periods.}
\label{fig:minipage3}       
\end{minipage}
\quad
\begin{minipage}[t]{0.49\linewidth}
 \includegraphics[width=\textwidth]{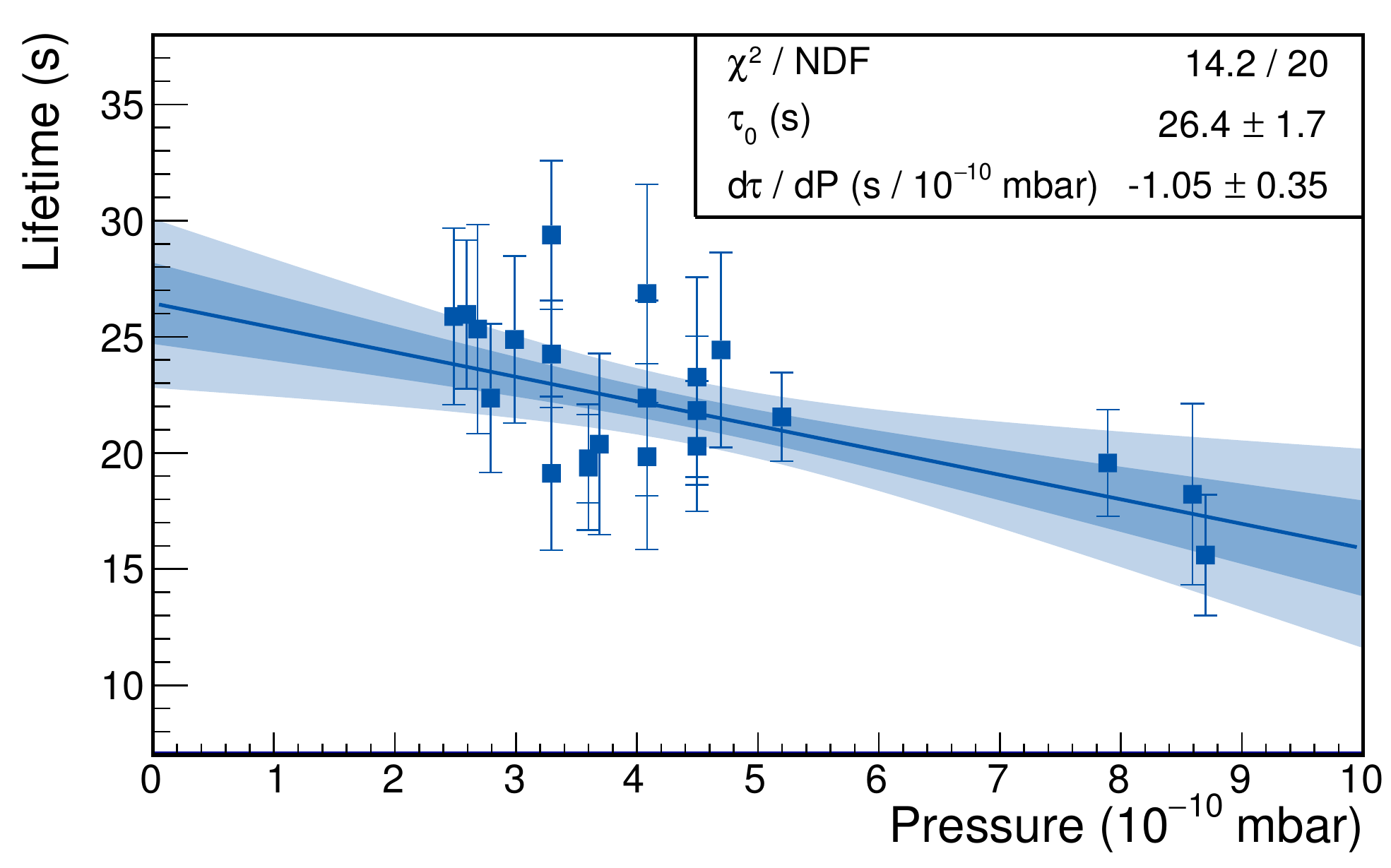}
\caption{Lifetime of the 5d$^2$D$_{5/2}$ state versus residual gas pressure in a single Ba$^+$ ion.
68$\%$ and 95$\%$ confidence intervals are given.}
\label{fig:minipage4}
\end{minipage}
\end{figure}
Collisions with background gas can reduce the lifetime of the metastable state.
In order to extrapolate the absolute value for the lifetime to zero pressure, the
lifetime $\tau_{D_{5/2}}$ was measured at different background pressures. Fig.~5 displays
the results for a selection of 1600 out of 5046 shelved periods. The uncertainty for the lifetime in
each value corresponds to the statistical error from fitting an exponential decay 
to the data. A range of pressures between $2.5 \times 10^{-10}$ and $8.7 \times 10^{-10}$ mbar was
explored by changing the temperature of the vacuum chamber in the range from 289
K to 296 K and by adjusting the pumping speed of the ion pump. For the small
change in temperature needed here, changes in the collision cross-sections between the ion and
the residual gas atoms can be neglected. A linear function is fitted to the data. 
The lifetime of the  5d$^2$D$_{5/2}$ state is found to be $\tau_{D_{5/2}} = 26.4(1.7)$~s. 
3446 shelved periods are used to check for systematics, such as potentially arising from laser intensities, 
laser frequency detunings, rf voltages for trap and effects from the operating conditions of the ion pump. 
No significant effects have been observed.
\begin{figure}[h]
\centering
\includegraphics[width=0.6\textwidth]{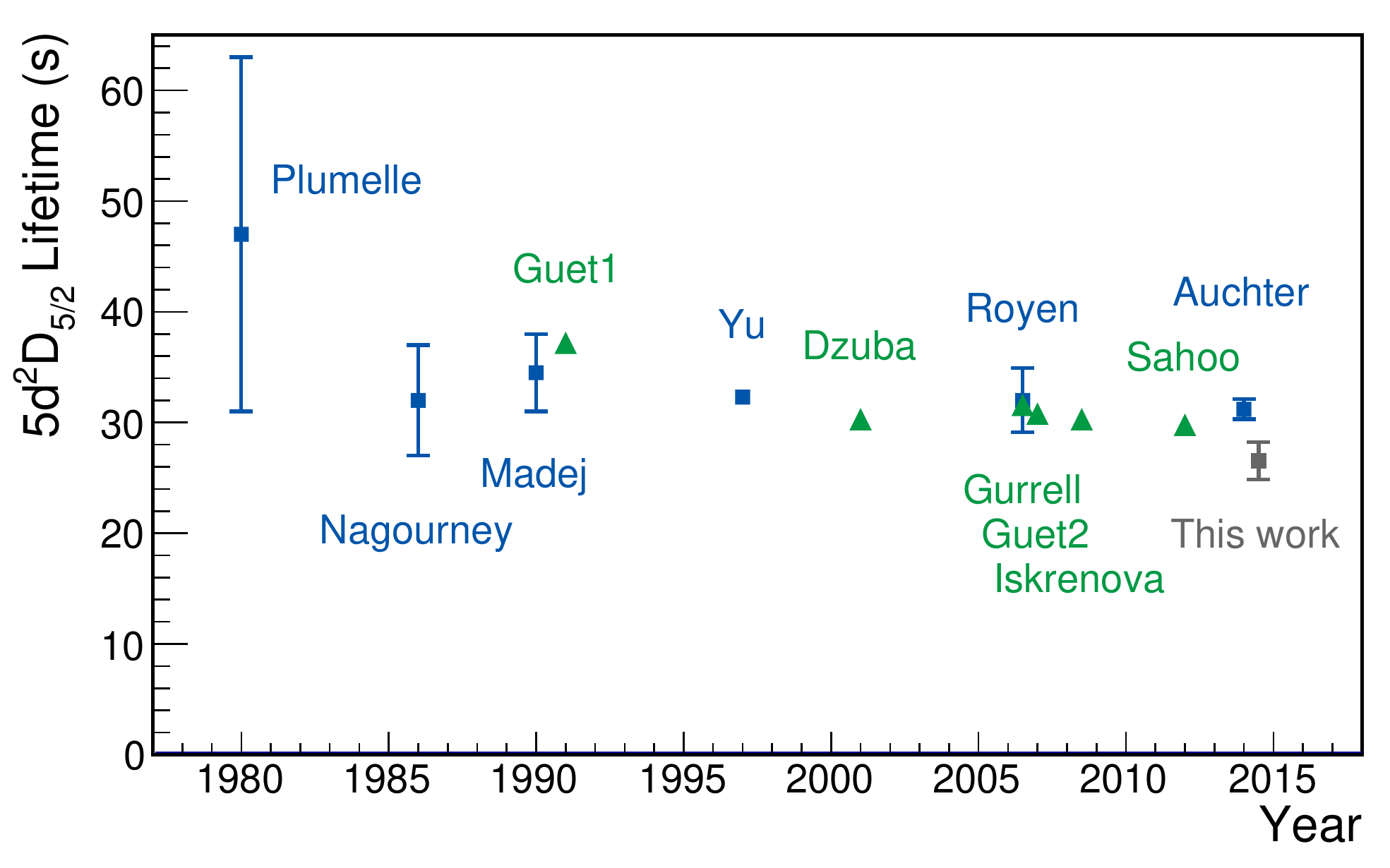}
 \caption{Lifetime of the 5d$^2$D$_{5/2}$ state in a single Ba$^+$ ion versus time within the last four decades. 
 Squares represent measurements and triangles represent calculated values. Note, the result Guet2 \cite{Guet1}
 differs from Guet1 \cite{Guet} by an omitted term.}
 \end{figure}
 \vspace*{-1.30cm}
\section{Conclusions}
\label{sec:4}
In summary, the lifetime of the metastable 5d$^2$D$_{5/2}$ has been measured for a single Ba$^+$ ion. 
The measured value is preliminary because cross checks for systematics are still ongoing.
Our result agrees within 2$\sigma$ with the most recent theoretical value $\tau_{D_{5/2}} = 29.8(3)$~s \cite{Sahoo2} 
and with the latest independent experimental value of $\tau_{D_{5/2}} = 31.2(9)$~s \cite{Auchter}. Fig.~6 displays the time 
evolution of the measurements and the theory values for the lifetime of the 5d$^2$D$_{5/2}$ state in a Ba$^+$ ion.
\vspace*{-0.50cm}
\begin{acknowledgements}
We thank Leo Huismann, Oliver B\"{o}ll, and Otto Dermois for their technical assistance. We acknowledge the 
financial support from FOM Programme 114 (TRI$\mu$P/AGOR) and FOM programme 125 (Broken Mirrors and Drifting Constants).
\end{acknowledgements}

\vspace{-0.50cm}

\end{document}